%% file: main.tex
\documentclass[10pt,twocolumn,letterpaper]{article}

\usepackage{cvpr}              %
\usepackage{makecell}

\input{preamble}
\definecolor{cvprblue}{rgb}{0.21,0.49,0.74}
\usepackage[pagebackref,breaklinks,colorlinks,allcolors=cvprblue]
{hyperref}
\usepackage{multirow}  %
\usepackage{multicol}
\usepackage{float}

\def\paperID{14506} %
\def\confName{CVPR}
\def\confYear{2025}

\newcommand{\inst}[1]{{$^{#1}$}}

\title{ReRAW: RGB-to-RAW Image Reconstruction via Stratified Sampling for Efficient Object Detection on the Edge}

\author{ 
    Radu Berdan\inst{1}\inst{!} \quad
    Beril Besbinar\inst{1}\inst{*} \quad
    Christoph Reinders\inst{2}\inst{*} \quad
    Junji Otsuka\inst{3} \quad
    Daisuke Iso\inst{1} \quad
    \\
    \inst{1}Sony AI \quad
    \inst{2}Leibniz University Hannover \quad
    \inst{3}Sony Group Corporation
    \\
    \inst{!}radu.berdan@sony.com \quad
    \inst{*}Equal Contribution. \quad
}

\begin{document}
\maketitle
\input{sections/0_Abstract}

\input{sections/1_Introduction}
\input{sections/2_Related_new}
\input{sections/3_Method}

\input{sections/4_Experiments}
\input{sections/5_Discussion}

\input{sections/6_Conclusion}

\newpage
{
    \small
    \bibliographystyle{ieeenat_fullname}
    \bibliography{main}
}

\end{document}

%% file: sections/0_Abstract.tex
\begin{abstract}
Edge-based computer vision models running on compact, resource-limited devices benefit greatly from using unprocessed, detail-rich RAW sensor data instead of processed RGB images. Training these models, however, necessitates large labeled RAW datasets, which are costly and often impractical to obtain. Thus, converting existing labeled RGB datasets into sensor-specific RAW images becomes crucial for effective model training. In this paper, we introduce ReRAW, an RGB-to-RAW conversion model that achieves state-of-the-art reconstruction performance across five diverse RAW datasets. This is accomplished through ReRAW’s novel multi-head architecture predicting RAW image candidates in gamma space. The performance is further boosted by a stratified sampling-based training data selection heuristic, which helps the model better reconstruct brighter RAW pixels. We finally demonstrate that pretraining compact models on a combination of high-quality synthetic RAW datasets (such as generated by ReRAW) and ground-truth RAW images for downstream tasks like object detection, outperforms both standard RGB pipelines, and RAW fine-tuning of RGB-pretrained models for the same task. The code is available at: \url{https://anonymous.4open.science/r/ReRAW-0C87/}
\end{abstract}

%% file: sections/1_Introduction.tex
\section{Introduction}
\label{sec:intro}

\begin{figure}[t]
    \centering
    \includegraphics[width=\columnwidth]{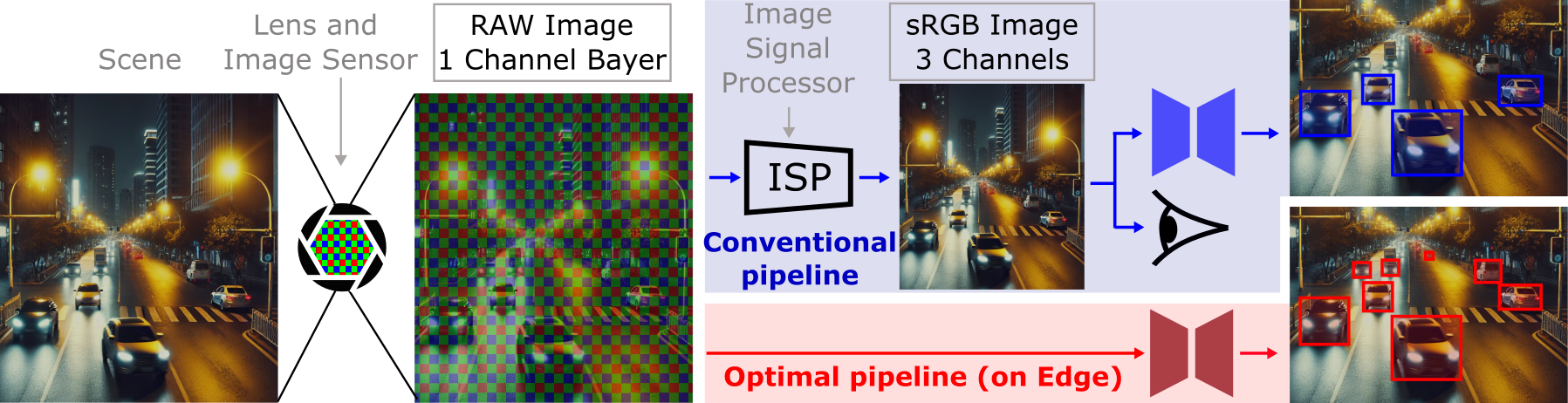}
    \caption{
    The conventional imaging pipeline involves an image sensor capturing a scene into a RAW image, converting that image into standard RGB fit for human consumption, and running computer vision tasks on these RGB images. The optimal pipeline would involve performing high-level tasks directly on the RAW images, on chip, and physically close to the image sensor.
    \vspace{-0.4cm}
    }
    \label{fig_1}
\end{figure}

The lifecycle of a digital image begins at the camera sensor, where incoming light from a scene is converted into electrical signals to form a RAW image -- a single-channel Bayer-pattern array \cite{bayer} where each pixel value corresponds linearly to the scene’s luminosity. These RAW images are then processed locally through a camera-specific Image Signal Processor (ISP), which applies multiple functions such as demosaicking, white balancing, tone mapping etc., to yield a compressed RGB image optimized for human perception, as shown in the conventional pipeline of Fig. \ref{fig_1}.

RGB images are preferred over RAW images due to their much smaller compressed size and compatibility with human vision, making them fast to transmit from device to device and convenient to store. However, unprocessed RAW images retain more detail, have a larger dynamic range (12-16 bits), and contain more information than their ISP-processed RGB outputs (see Supplementary \ref{sec:s-raw-info}). This suggests that all else being equal, computer vision models such as object detectors trained directly on RAW data could outperform those trained on RGB. Hence, when operating in resource-constrained environments such as on the edge, the reduction in model accuracy due to limited compute can be mitigated by utilizing a richer signal source such as RAW images, compared to RGB images. Additionally, bypassing the ISP improves power consumption and speed, as in the optimised pipeline in Fig. \ref{fig_1}.

Nevertheless, RGB pre-trained object detectors perform suboptimally on RAW images due to the domain gap. 
Several approaches have attempted to bridge this gap with learnable adapters~\cite{adaptor-rod, adaptor-genisp} or other non-linear scaling functions beyond standard ISP processing~\cite{adaptor-cooked, learnable-1} as well as adopting traditional feature extraction methods, such as Histogram of Oriented Gradients (HOG)~\cite{adaptor-hog}.
Although end-to-end training of object detectors with these adaptations yields promising results, performance is still limited by the additional computational burden and the scarcity of labeled RAW datasets.
Alternatively, a reverse ISP function that converts RGB images back to RAW sensor output could leverage widely available labeled RGB datasets.
Recent methods have explored learning reverse ISP functions~\cite{invisp, upi, paramisp} or adopting generative models such as CycleGAN ~\cite{EfficientVisualComputing2024}
or diffusion models ~\cite{rawdiffusion}.
However, capturing the true color profile or details in bright regions in RAW images, particularly for dark scenes, remains a challenge.

Hence, in this work, we introduce ReRAW, a multi-head reverse ISP model designed to reconstruct sensor-specific RAW images from RGB inputs, without RAW meta-data, faithfully capturing the color characteristics of a target camera sensor. 
ReRAW enables the conversion of large RGB datasets into realistic camera-specific RAW to train object detectors that are mainly targeted for edge deployment. 
Our design improves reconstruction accuracy for both day and night images, effectively reproducing over-exposed regions in dark RAW images, such as those captured at night.
Our primary contributions are as follows:

\begin{itemize} 
    \item We propose a novel reverse ISP model, ReRAW, capable of reconstructing sensor-specific RAW images from RGB with high fidelity. The model employs a multi-head ensemble architecture that generates multiple gamma-corrected RAW images, subsequently scaled and combined to match the sensor characteristics. 
    \item A stratified sampling technique for training data, which results in the trained ReRAW to better capture bright regions in the converted RAW images, compared to using full training datasets or random sampling. Adding a logarithm-based loss function further boosts performance achieving state-of-the-art conversion accuracy compared to competing methods, on five diverse RAW datasets.
    \item Empirical evidence that pretraining object detectors on high-quality synthetic RAW datasets produced by ReRAW, followed by fine-tuning on real RAW data for specific tasks, outperforms models fine-tuned from an RGB-pretrained baseline. This approach removes the need for ISPs in traditional imaging pipelines on the edge and eliminates the need for extra fixed or learnable adapters to align RAW data with RGB-pretrained models. Our direct RAW training from scratch is effective, provided the synthetic RAW images are of high quality.
\end{itemize}

%% file: sections/2_Related_new.tex
\section{Related Work}
\label{sec:related}

\subsection{RAW Images}
RAW images, with their higher dynamic range and linear noise profile, offer advantages over standard RGB images, especially in low-light conditions. 
However, they are usually $5-10\times$ larger in size than compressed RGB, and large RAW datasets ($>100K$ images), to the best of our knowledge, do not exist. 
Nonetheless, with limited data, recent studies have demonstrated improved outcomes in image classification~\cite{maxwell2023log,maxwell2024logarithmic}, object detection~\cite{xu2023toward, yoshimura2023dynamicisp,ljungbergh2023raw,wang2024adaptiveisp,dutta2022seeing,cui2025raw}, semantic segmentation~\cite{cui2025raw}, and instance segmentation~\cite{chen2023instance} with models designed for the RAW domain.

\subsection{RGB-to-RAW Reconstruction}
The advantages of RAW images, coupled with the extremely limited availability of RAW datasets, have fueled interest in reconstructing RAW images from RGB counterparts to expand labeled datasets. 
Traditional methods determine the relationship between a camera’s output intensity and the incident light by capturing multiple images at various controlled exposure levels, with varying levels of complexity \cite{mitsunaga1999radiometric, grossberg2003determining, lin2005determining, debevec2023recovering, chakrabarti2009empirical, kim2012new, chakrabarti2014modeling}. 
However, these approaches require calibration for each camera, necessitating multiple parameterized models for different settings. 
In contrast, modern data-driven algorithms \cite{afifi2021cie, brooks2019unprocessing, conde2022model, condeReversedImageSignal2022, gharbi2016deep, schwartz2018deepisp, zamir2020cycleisp, xingInvertibleImageSignal2021, liang2021cameranet, liu2022deep} leverage advanced machine learning to address this complex inverse problem without calibration. 
A common strategy involves simulating single or groups of ISP functions with neural networks~\cite{liang2021cameranet, liu2022deep, yoshimura2023dynamicisp}, which requires sensor-specific configuration and training while limiting the flexibility of a data-driven perspective. 
Alternatively, other methods encapsulate ISP functions within a single network~\cite{zamir2020cycleisp, xingInvertibleImageSignal2021, brooks2019unprocessing} yet they demand extensive RGB-RAW paired datasets.

\begin{figure*}
    \centering
    \includegraphics[width=\textwidth]{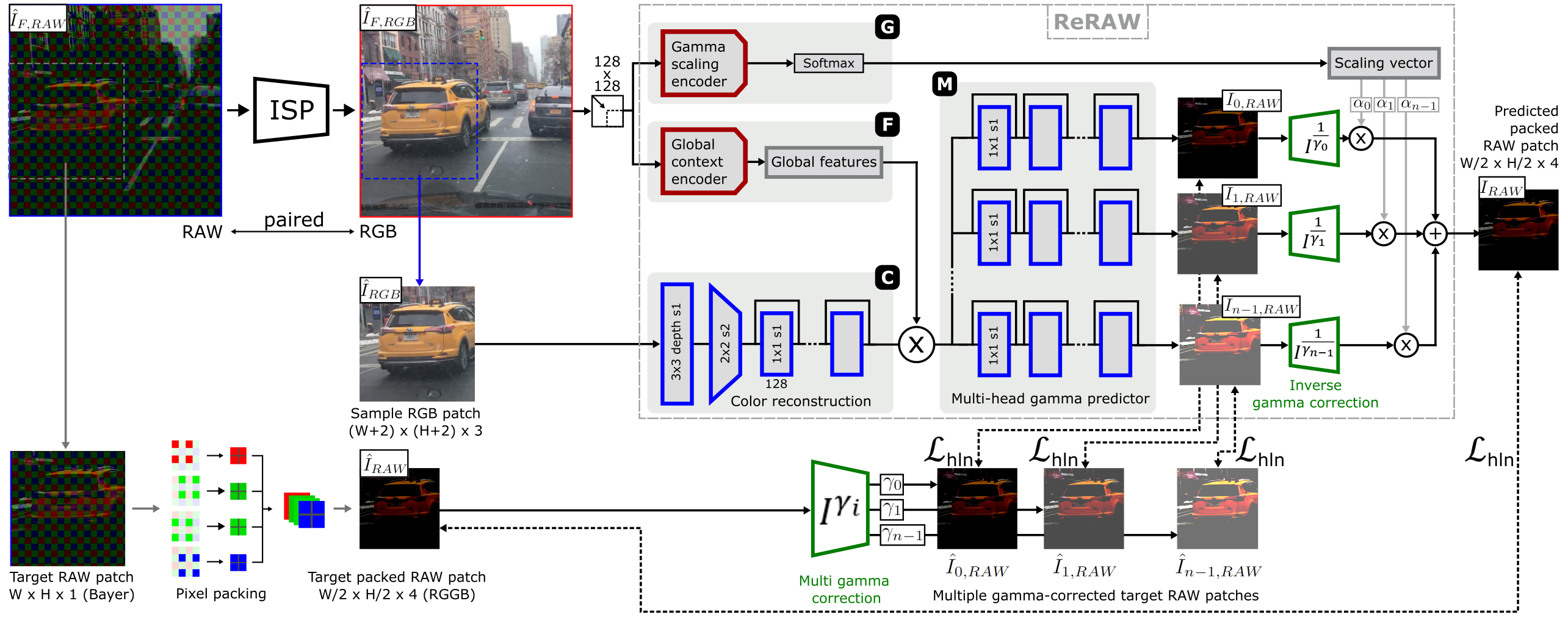}
    \caption{
    Illustration of ReRAW architecture and training data flow. A Global Context Encoder (F) extracts features from the full RGB image to guide the Color Reconstruction network (C), while a Multi-head Gamma Predictor (M) generates multiple gamma-corrected RAW patches. These patches are then degammaed (inverse gamma correction), scaled by a scaling vector, predicted by a Gamma Scaling Encoder (G) from the original RGB image, and summed to form the final RAW patch. Losses are applied between each intermediate gamma-corrected RAW patch and target, as well as between the final RAW output and target RAW.}
    \label{fig_3}
    \vspace{-0.2cm}
\end{figure*}

\subsection{Object Detection on the Edge}
Modern deep learning techniques have significantly advanced object detection performance~\cite{girshick2014rich,lin2017feature,redmon2016you,tian2019fcos,redmon2016you,renFasterRCNNRealtime2015,liu2023grounding}.
Beyond algorithmic advancements, this improvement could also be attributed to the increasing scale of detection models and the availability of large labeled datasets. 
However, many practical object detection applications operate at the edge, where limited computational power, memory, and a restricted power budget are common constraints. 
These requirements can be addressed either by compressing large models through knowledge distillation~\cite{li2023object}, quantization~\cite{li2019fully}, and/or pruning~\cite{figurnov2017spatially,liang2022edge} or by designing lightweight models from scratch~\cite{howard2017mobilenets, sandler2018mobilenetv2,howard2019searching,zhang2018shufflenet,li2021micronet,tan2019efficientnet}.

We hypothesize that when edge computing is paired with RAW sensor data, the resulting edge-based imaging and sensing systems can achieve greater versatility in monitoring challenging scenes and improved overall performance. 

%% file: sections/3_Method.tex
\section{ReRAW}
\label{sec:reraw}
Motivated by the lack of large-scale labeled RAW datasets, and analysing the shortcomings of previous reverse ISPs, we design ReRAW as a universal RGB-to-RAW converter that can handle both daytime and nighttime images, with mild or strongly skewed RAW pixel distributions. 
\subsection{Overview}
ReRAW is designed to reconstruct a $W/2\times H/2\times4$ packed RGGB (RAW) image patch ($\hat{I}_{\text{RAW}}$), given both an input $W\times H\times3$ RGB image patch ($\hat{I}_{\text{RGB}}$) and the full RGB image ($\hat{I}_{\text{F,RGB}}$) from where the RGB patch originates from. We convolve ReRAW over an input RGB image to reconstruct the full required RAW image ($\hat{I}_{\text{F,RAW}}$).

The model's unique feature is its prediction in gamma space, over multiple gamma candidates, via a multi-head architecture. Gamma-corrected patch candidates are re-linearised (by applying an inverse gamma process) and proportionally averaged by a weight vector predicted by a Gamma Scaling Encoder from the original full RGB image. In this way, the model learns to select input image-dependent gamma transformations that would facilitate a better RAW conversion. Additionally, training via a stratified sampling data selection technique helps in mitigating the extreme skew of pixel values commonly found in RAW images.

\subsection{Architecture}
The full network architecture is shown in Fig.~\ref{fig_3}. The model consists of a Color Reconstruction Network (C), a Global Context Encoder (F), a Multi-head Gamma Predictor (M), and a Gamma Scaling Encoder (G). The model is trained to predict $n$ gamma-corrected RAW target patches ($I_{\text{i,RAW}}$) from an input RGB patch and its container RGB image:

\begin{equation}
    \{I_{i,RAW}\}_{i=0}^{n-1} = M(C(\hat{I}_{\text{RGB}})\times F(\hat{I}_{\text{F,RGB}})) 
    \label{eq:network-1}
\end{equation}

The gamma-corrected patch candidates $I_{i,RAW}$ are then degammaed (inverse gamma process to re-linearize the images) and each multiplied by a scaling factor predicted by the Gamma Scaling Encoder from the original full RGB image. The linearised and scaled RAW candidate patch images are then summed to output the final RAW patch $I_{\text{RAW}}$.

\paragraph{Color Reconstruction Network (C)}
This module consists of an initial depth-wise convolutional layer with kernel size $3\times3$, stride 1, 3 groups, and 96 channels. It is then followed by a $2\times2$ stride 2 convolutional layer with 128 channels. This reduces the input spatial dimension of the RGB patch from $(W+2)\times (H+2)$ (not using padding) to $(W/2)\times (H/2)$. The output is then fed into a residual network consisting of 8 point-wise convolutional layers with depth 128. The output of the $C$ network is thus a latent tensor of $(W/2)\times (H/2)\times128$, where each spatial multichannel pixel has a receptive field size of only $4\times4$ in the original RGB patch.

\paragraph{Global Context Encoder (F)}
This module encodes general characteristics from the original RGB image (scaled to $128\times128$) such as luminosity and color space features, and uses this information to modulate the RGB-to-RAW color conversion.
The module consists of a ResNet18~\cite{resnet} where the last classification layer has been replaced with a linear layer of $1\times1\times128$ output size. This output tensor is then expanded to shape $(W/2)\times(H/2)\times128$ and multiplied with the output tensor of the Color Reconstruction Network (C). We found empirically that multiplication gave better results than addition or concatenation.

\paragraph{Multi-head Gamma Predictor (M)}
This network holds $n=10$ parallel heads, each consisting of 8 residual point-wise convolutional layers with depth 128, and output depth 4. Each head outputs a candidate gamma-corrected RAW patch ${I}_{\text{i,RAW}}$. The motivation for using a multi-head approach is that converting to RAW in gamma space can be helpful when there are significant differences between the input RGB and output RAW pixel distributions \cite{multigamma}. For instance, daytime datasets with minimal ISP adjustments typically present an easier reconstruction task, whereas nighttime datasets, which necessitate more extensive ISP operations, pose a more challenging reconstruction problem. To address these varying complexities, the multi-head strategy is designed to learn distinct transformation pathways.

\paragraph{Gamma Scaling Encoder (G)}
The Gamma Scaling Encoder, also a ResNet18, learns to encode the full RGB image into a scaling vector of softmax-normalized values of size $n$: 

\begin{equation}
    \{\alpha_{i}\}_{i=0}^{n-1} = G(\hat{I}_{\text{F,RGB}}),  \sum_{i=0}^{n-1} \alpha_{i} = 1
    \label{eq:network-2}
\end{equation}

The output gamma-corrected RAW patches (${I}_{\text{i,RAW}}$) from the Multi-head Gamma Predictor (G) are then re-linearized (de-gammaed), scaled by each $\alpha_{\text{i}}$ value from the dynamic scaling vector and summed in order to output the final RAW predicted patch ${I}_{\text{RAW}}$:

\begin{equation}
    I_{RAW} = \sum_{i=0}^{n-1} I_{i,RAW}^{\frac{1}{\gamma_{i}}}\times\alpha_{i}
    \label{eq:network-4}
\end{equation}

\begin{figure*}
  \centering
  \begin{subfigure}{0.63\textwidth}
    \includegraphics[width=\linewidth]{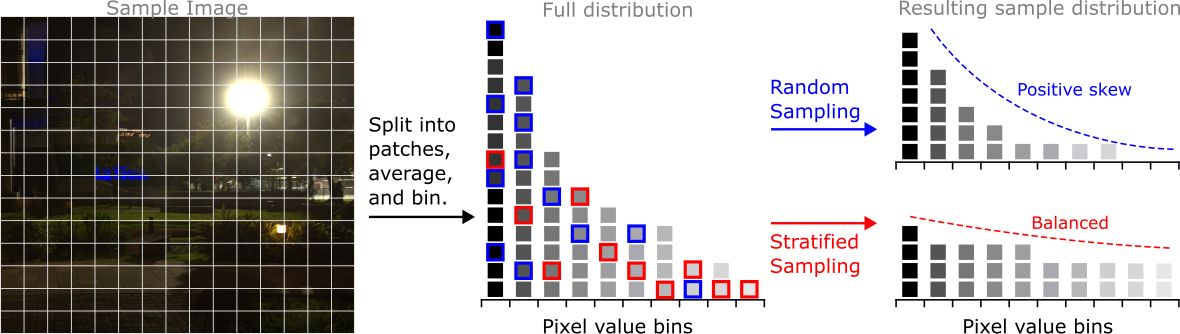}
    \caption{Stratified vs random sampling illustration.}
    \label{fig:stratified-vs-random-a}
  \end{subfigure}
  \hfill
  \begin{subfigure}{0.25\textwidth}
    \includegraphics[width=\linewidth]{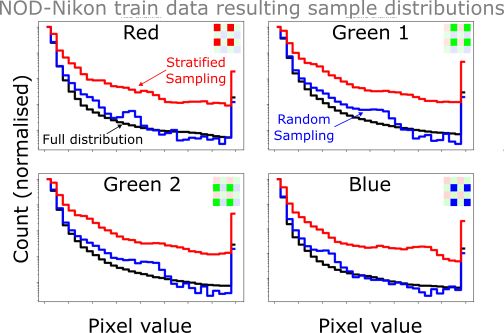}
    \caption{Sampling on NOD-Nikon.}
    \label{fig:stratified-vs-random-b}
  \end{subfigure}
  \begin{subfigure}{0.1\textwidth}
    \includegraphics[width=\linewidth]{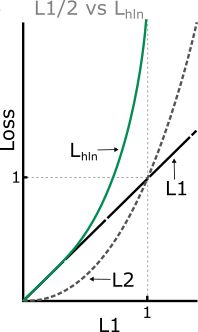}
    \caption{Losses}
    \label{fig:stratified-vs-random-c}
  \end{subfigure}
  \caption{
    (a) Comparison of random vs stratified sampling for RGB-to-RAW conversion training data preparation. As explained in \ref{sec:stratified}, random sampling results in a pixel intensity distribution similar to the original image, while stratified sampling results in a much more balanced pixel intensity distribution. (b) Histograms show pixel value distributions for each color channel from patches of NOD-Nikon RAW dataset for random (red), full (black) or stratified (blue) sampling methods showing the desired effect. (c) Comparisons of losses. 
    }
  \vspace{-0.2cm}
  \label{stratified_sampling}
\end{figure*}

\subsection{Training Objective}
The network is optimized to predict a RAW image patch $I_{\text{RAW}}$, given an input RGB patch $\hat{I}_{\text{RGB}}$ and the corresponding full RGB image 
$\hat{I}_{\text{F,RGB}}$. The network outputs $n$ intermediate gamma-corrected RAW patches $I_{\text{i,RAW}}$, and a final RAW patch $I_{\text{RAW}}$. In low-level image processing, training objectives normally comprise of minimising the $L_{\text{1}}$ or $L_{\text{2}}$ distance between a target and a predicted image ~\cite{l1l2loss}. Further, in order to address the data distribution skew towards lower pixel values, a logarithm-based loss function is sometimes used ~\cite{lnloss}. We design our own loss function \textit{hard-log} ($\mathcal{L}_{\text{hln}}$) which heavily penalizes wrongly reconstructed pixel values, whilst converging to a $L_{\text{1}}$ loss for lower error pixels (Fig. ~\ref{fig:stratified-vs-random-c}). This loss function helps better reconstructing sparse bright pixels in mostly dark images.
\begin{equation}
    \mathcal{L}_{hln}(\hat{I}, I) = \frac{-1}{CHW} \lVert(ln(1-|\hat{I}-I| + \epsilon)\rVert_{1}
    \label{eq:loss}
\end{equation}

For the Multi-Head Gamma Predictor we chose 10 gamma values $\gamma_{\text{i}} \in \{0.1, 0.2, ... 1\}$. The target gamma-corrected candidate RAW patches are therefore:

\begin{equation}
    \hat{I}_{i,RAW} = \hat{I}_{RAW}^{\gamma_{i}} 
    \label{eq:network-3}
\end{equation}

The overall training objecting is thus minimizing the loss both between the candidates ($I_{\text{i,RAW}}$) and target gamma-corrected patches ($\hat{I}_{\text{i,RAW}}$), and between the final RAW patch ($I_{\text{RAW}}$) and target RAW patch ($\hat{I}_{\text{RAW}}$):

\begin{equation}
    \mathcal{L} = \mathcal{L}_{hln}(\hat{I}_{RAW}, I_{RAW}) + \sum_{i=0}^{n-1} \mathcal{L}_{hln}(\hat{I}_{i,RAW}, I_{i,RAW}) 
    \label{eq:loss_g}
\end{equation}

\subsection{Stratified Sampling}
\label{sec:stratified}

Training ReRAW needs paired RGB and RAW images. These can be created by capturing RAW images with a camera and using the camera's ISP or a generic ISP to produce the paired RGB images. The training set is made up of small image patches taken from each RGB image, along with the matching patch from the paired RAW image. The network takes one RGB patch and the full RGB image it came from as inputs.

The pixel distribution in RAW images is in usual cases (natural scenes) skewed toward darker values, and especially in nighttime images. Examining pixel distributions from all patches in a dataset shows a bias toward low-intensity values. This bias remains even when sampling a subset of patches randomly from the RAW image dataset. Training an RGB-to-RAW converter would therefore tend to prioritize reconstructing darker pixels over brighter regions, which may contain important information useful for other high-level computer vision tasks.

To address this bias, we propose a \textit{stratified sampling} technique to create a paired RAW to RGB patch dataset that better balances the pixel distributions in both domains. This aims to improve reconstruction performance, especially in the brighter regions of the image. The process and its impact on pixel distribution are shown in Fig.~\ref{fig:stratified-vs-random-a}. The steps for the stratified sampling method are listed below:

\begin{enumerate} 
    \setlength\itemsep{0pt} 
    \item Split each RGB image in the dataset into patches and compute the average brightness for each color channel. 
    \item Bin the patches based on their average brightness for each channel, resulting in three vectors of binned patches. We use 10 bins: $[0 - 0.1), ..., [0.9 - 1.0)$.
    \item Uniformly select one bin, then uniformly pick an RGB patch from that bin and its corresponding RAW patch. 
\end{enumerate}

The above steps describe our method for selecting an RGB-RAW patch pair from the paired images. We repeat this process multiple times for each color channel and each image in the dataset to build our training set. Compared to random sampling, the stratified sampling method results in a more even pixel intensity distribution, as shown in Fig.~\ref{fig:stratified-vs-random-b}. While it is not possible to achieve a perfectly uniform distribution (as we sample full 4-channel pixel values, not individual channels), this method significantly improves uniformity.

%% file: sections/4_Experiments.tex
\section{Experiments}
\label{sec:experiments}
\input{tables/table-revisp}   
We perform extensive experiments analyzing the performance of our RGB-to-RAW reconstruction network ReRAW. We benchmark our network on several challenging RAW datasets (daytime and nighttime) and against competing methods and show state-of-the-art performance. Further, we apply ReRAW to convert large labeled RGB datasets to sensor-specific RAW, and show that training RAW object detectors (OD) from scratch on combined synthetic and ground-truth labeled RAW data outperforms traditional RGB data pretraining and finetuning. We demonstrate this via a strict 1-to-1 comparison, over three different object detectors and over two different RAW/RGB datasets (one day and one nighttime) to show the validity of our training recipe.

\subsection{Datasets}
\label{sec:datasets}
\paragraph{RGB-to-RAW conversion}
We utilize five different RAW datasets. From the MIT-Adobe FiveK \cite{fivek} collection we select images taken with the Nikon D700 and Canon EOS 5D SLR cameras to create two datasets of 542 and 707 RAW images, respectively. We name these: FiveK-Nikon and FiveK-Canon. We also utilize the NOD \cite{noddataset} RAW nighttime dataset consisting of two sets of images captured by a Nikon D750 and Sony RX100 VII SLR cameras. We named these as NOD-Nikon and NOD-Sony, each consisting of 4.0k and 3.8k images, respectively. Finally, we utilise PASCALRAW \cite{pascalrawdataset}, comprising of 4.2k RAW images captured by a Nikon D3200 DSLR camera. 

For all RAW datasets, we use $rawpy$ \cite{rawpy} to convert the RAW files to RGB images to create RAW-RGB pairs at full resolution. We use a 80/20 train/test split.

\paragraph{Object Detection}
We utilise PASCALRAW and NOD-Nikon RAW OD datasets to benchmark our models trained on synthetic RAW images. PASCALRAW contains mostly daytime images while NOD-Nikon contain strictly nighttime images. Both datasets are labeled with objects of 3 classes (person, car and bicycle). 

To create our large synthetic labeled RAW datasets, we utilise the BDD100K ~\cite{bdd100k} autonomous driving OD dataset. We select only images that contain at least one of the 3 classes of interest, and further split this into daytime and nightime images, using the provided image meta-labels. We extract a 3 class OD \textit{daytime subset} of 36.5k images with 476k instances, and a 3 class OD \textit{nighttime subset} of 27.5k images and 263k annotations. 

Utilising PASCALRAW, the daytime RAW OD dataset, and the daytime BDD subset, we create three variations:
\begin{enumerate}
  \item \textbf{BDD-RGB}: contains a mix of daytime BDD RGB images and the ground truth RGB images from the PASCALRAW train split.
  \item \textbf{BDD-ReRAW-R}: contains daytime BDD RGB images converted to synthetic RAW by the ReRAW model trained on PASCALRAW via \textit{random sampling} patch selection (ReRAW-R), combined with the ground truth RAW images from the PASCALRAW train split.
  \item \textbf{BDD-ReRAW-S}: contains daytime BDD RGB images converted to synthetic RAW by the ReRAW model trained on PASCALRAW via \textit{stratified sampling} patch selection (ReRAW-S), combined with the ground truth RAW images from the PASCALRAW train split.
\end{enumerate}
Utilising NOD-Nikon, we prepare the same 3 variations of datasets, however using the nighttime BDD subset, and the ReRAW variations trained on NOD-Nikon. A visualisation of the converted images is shown in Supplementary Fig. \ref{bdd_reraw}.

\begin{figure*}
    \centering
    \includegraphics[width=\textwidth]{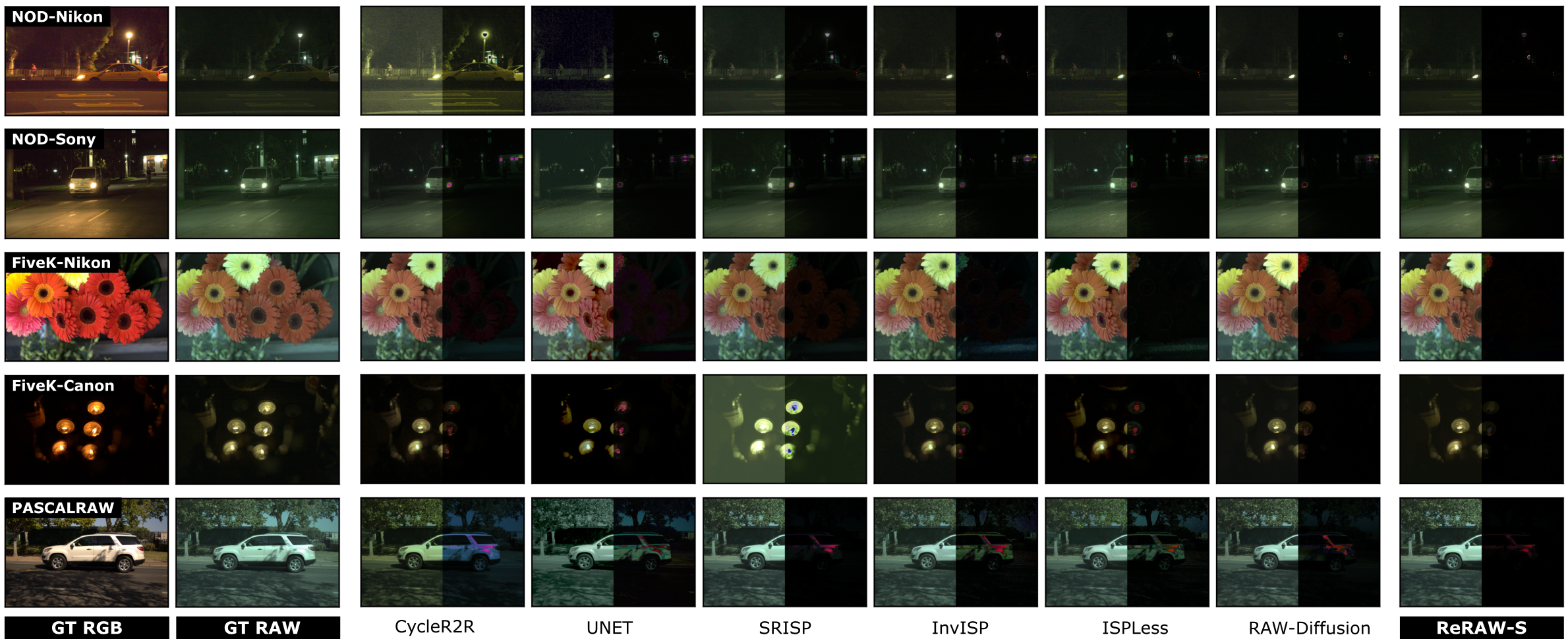}
    \caption{
    Qualitative comparison of RAW reconstruction for several competing reverse ISP models and ReRAW-S. First two columns show ground truth RGB followed by ground truth RAW. For each model, the input is formed just from the ground truth RGB image. Each row shows one example from each dataset. Each reconstructed image is split in half, where the left half shows the gamma-corrected reconstructed image, and right half shows an error map vs ground truth RAW image. Best seen in color. }
    \label{fig_4}
    \vspace{-0.3cm}
\end{figure*}
\subsection{RGB-to-RAW Reconstruction}

\paragraph{Training Setup}
We sample about six $68\times68$ RGB patches and their corresponding $32\times32$ RAW patches per image pair, once randomly and once using our stratified sampling method to create two separate training subsets per RGB-RAW dataset. We use each of these to train a separate ReRAW variant, ReRAW-R utilising the random sampled subset and ReRAW-S utilising the stratified sampled subset. Each target RAW patch is black-level subtracted and max-normalised. The full context RGB image is downscaled to $128\times128$ and randomly cropped to 0.9 of its area for each patch sample.
We train each ReRAW model using the Adam optimizer \cite{adam} with a batch size of 32. Training is done for 128 epochs using cosine annealing with warm restarts every 16 epochs, with a starting learning rate of $10^{-3}$ and decaying to $10^{-5}$.

\paragraph{Evaluation}
For each trained model we convert the RGB test set for each RGB-RAW datasets into their synthetic RAW counterparts at full resolution and evaluate both PSNR and SSIM \cite{ssim} (Structural Similarity Index Masure) compared to the original RAW images. We average PSNR and SSIM for all reconstructed images in each dataset and report the results. 

\paragraph{Results}
We compare ReRAW with several state-of-the-art reverse ISPs: CycleR2R \cite{EfficientVisualComputing2024}, a CycleGAN based method, UNet \cite{conde2022reversed}, SRISP \cite{FAL-revisp}, InvISP \cite{invisp}, ISPLess \cite{adaptor-cooked}, and RAW-Diffusion \cite{rawdiffusion}, a diffusion-based method. For SRISP, we use the mean global feature of all training images as test-time reference. For InvISP and ISPLess, we also include variants where during the inverse process, the ground truth RGB image is used as input, denoted as InvISP$^{+}$ and ISPLess$^{+}$. The conversion results on all the five RAW datasets listed before are shown in Table \ref{tab:reraw}.
ReRAW outperforms all listed reverse ISPs in terms of PSNR and SSIM. RAW-Diffusion achieves second best overall results. We report two variations of our model: ReRAW-R - trained with a dataset of patches selected randomly; and ReRAW-S - trained on a dataset of patches selected via our stratified sampling method. ReRAW-S achieves the highest reconstruction PSNR for NOD-Nikon, NOD-Sony, FiveK-Canon, and PASCALRAW datasets. Also, ReRAW-S, enabled by the stratified sampling technique, reconstructs the high intensity RAW pixel values better compared to ReRAW-R, as can be seen in Supplementary Fig. \ref{stratified_vs_random}.

Figure \ref{fig_4} shows qualitative results of synthetic RAW images reconstructed from original RGB images using our models and competing reverse ISPs, compared to their ground truth RAWs. Due to both the stratified sampling technique and logarithm-based loss function, highlighted regions of the RAW images are better reconstructed by ReRAW-S compared to competing methods. This better reconstruction can also be visualised when plotting the ground truth RAW pixel values vs synthetic RAW pixel values for all RAW color channels, as shown in Supplementary \ref{graphs}. ReRAW-S achieves a more linear relationship between predicted and real pixel values, boosting a higher PSNR.

\input{tables/table-ablation-loss}  
\input{tables/table-ablation-architecture}

\paragraph{Ablation Study}
We perform experiments to study how each component and learning heuristic impacts the performance of ReRAW.

We test the impact of different loss functions on training ReRAW, as shown in Table \ref{tab:ablation-loss}. Using the stratified-sampling training set, we train the model with $\mathcal{L}_{1}$, $\mathcal{L}_{2}$, and our proposed $\mathcal{L}_{hln}$ loss functions. The logarithm-based loss performs best across datasets, likely because it penalizes poorly reconstructed high-value pixels more strongly than $\mathcal{L}_{1}$ and $\mathcal{L}_{2}$, leading to better overall conversion performance.

We also ablate modules and modify network hyperparameters, with the results shown in Table \ref{tab:ablation-architecture}. The context encoder proves important for PSNR performance, since it proves global color modulation parameters can be extracted from the converted RGB scene. Additionally, the multi-head architecture allows the model to convert various gamma-corrected patches and select the best ones. 
\input{tables/table-mAP}

\subsection{Object Detection}
\paragraph{Training Setup}
We evaluate several recipes of training small object detectors for running on the edge. We train 3 different single-stage object detectors: RTMDet-s \cite{rtmdet}, YOLO-X-s \cite{yolox}, and SSD with a MobileNet-v2 backbone \cite{MobileNet}, each designed for efficient, real-time object detection through simplified architectures that optimize inference speed and accuracy.
We train each detector in two stages: a pretraining stage on variations of a large custom dataset extracted from BDD100K, as listed in Section \ref{sec:datasets}, and a finetuning stage on small real-world datasets of interest. Four different pretraining/finetuning combinations are tested in order to fairly evaluate the differences in performance between RGB and RAW trained OD models, and on two datasets: PASCALRAW (daytime) and NOD-Nikon (nightime).
We pretrain each detector from scratch for 50 epochs on the full custom BDD dataset (BDD-RGB or BDD-ReRAW-R/S), using stochastic gradient descent (SGD) with a cosine annealing schedule. Base learning rates are $0.001$ for RTMDet, $0.002$ for YOLOX and $0.015$ for SSD, decaying to $0.1\times$. Random flip, scale and mosaic are applied as augmentations only during pretraining. Each detector was then finetuned for 8 epochs on the ground-truth train set of PASCALRAW or NOD-Nikon, with starting learning rate of $0.1\times$ of base and decaying to $0.01\times$, also on a cosine annealing schedule. We keep the training heuristics identical per detector in order to allow a fair 1-to-1 comparison between each training dataset combinations.

\paragraph{Evaluation}
Each trained detector was evaluated on the PASCALRAW and NOD-Nikon RGB or RAW test sets, depending on the modality of the finetuning set, on mean Average Precision (mAP, mAP50, and mAP75).

\paragraph{Results}
The object detection training results are shown in Table \ref{tab:od}. The results under the PASCALRAW columns have been obtained using the daytime BDD-RGB and BDD-ReRAW datasets, and for the NOD-Nikon column, using the nighttime sets (as explained in Section \ref{sec:datasets}). 

The line a) result for each detector represents the baseline traditional pipeline of RGB pretraining and finetuning. Line b) involves taking the RGB pretrained model and finetuning it on RAW images (with the 2 green channels averaged). Although this is an immediate solution for adapting foundation RGB models to RAW, this method generally underperforms due to the domain gap. For lines c) and d), the detectors have been pretrained on synthetic large RAW image datasets converted from BDD-RGB using ReRAW-R (c) and ReRAW-S (d), and finetuned on real RAW images.

Training heuristic d) always outperforms the traditional RGB pipeline a), on a 1-to-1 comparison, for multiple detectors and on both a daytime, and a more difficult nighttime dataset. Additionally, training heuristic c) underperforms both d) and a) (for YOLOX and SSD), due to the large synthetic pretraining RAW dataset being lower fidelity (PSNR) than d). This underscores that the quality of the reverse ISP used to generate the large synthetic RAW pretraining dataset is important. The second performing training heuristic for the daytime dataset (PASCALRAW) is the RGB pipeline a). This is because the pixel distribution of daytime RGB images is closer to their original RAW versions. In contrast, for the nighttime dataset (NOD-Nikon), the second-best heuristic is b) or c), showing that in low-light conditions, RAW is optimal.

%% file: tables/table-revisp.tex
\begin{table*}[t]
  \centering
  \begin{tabular}{l|cc|cc|cc|cc|cc}
    \toprule
    Dataset & 
        \multicolumn{2}{c}{\textbf{NOD - Nikon}} & 
        \multicolumn{2}{c}{\textbf{NOD - Sony}} &
        \multicolumn{2}{c}{\textbf{FIVEK - Nikon}} &
        \multicolumn{2}{c}{\textbf{FIVEK - Canon}} &
        \multicolumn{2}{c}{\textbf{PASCALRAW}} \\
    Metric & PSNR & SSIM & PSNR & SSIM & PSNR & SSIM & PSNR & SSIM & PSNR & SSIM \\
    \midrule
    CycleR2R \cite{EfficientVisualComputing2024} & 
        24.51 & 0.5805 & 
        22.06 & 0.5069 & 
        24.60 &	0.8768 & 
        25.28 & 0.8582 & 
        26.65 & 0.7785 \\
    UNet \cite{conde2022reversed}& 
        34.58 & 0.9279 & 
        34.93 & 0.9067 & 
        27.18 &	0.8882 & 
        25.83 & 0.8895 & 
        27.81 & 0.8831 \\  
    SRISP \cite{FAL-revisp} &
        35.04 & 0.8953 &
        33.89 & 0.8628 &
        24.28 & 0.8313 &
        26.34 & 0.8164 &
        31.79 & 0.9490 \\
    InvISP \cite{invisp} &
        27.98 & 0.8843 &	
        28.37 & 0.8764 &
        27.09 & 0.9142 &	
        23.81 & 0.8596 &
        27.34 & 0.9120 \\  
    InvISP$^{+}$ \cite{invisp} &
        37.20 & 0.9708 &
        35.86 & 0.9499 &
        26.41 & 0.9093 &
        26.61 & 0.8995 &
        31.07 & 0.9507 \\    
    ISPLess \cite{adaptor-cooked} &    
        27.27 & 0.8867 &	
        27.18 & 0.8714 &
        27.79 & 0.9173 &	
        24.86 & 0.8728 &
        26.30 & 0.9057 \\
    ISPLess$^{+}$ \cite{adaptor-cooked} &
        37.14 & 0.9688 &	
        35.69 & 0.9489 &
        27.88 & 0.9093 &	
        26.85 & 0.8898 &
        30.45 & 0.9336 \\
    RAW-Diffusion \cite{rawdiffusion} &
        39.82 & 0.9804 & 	
        38.22 & 0.9658 & 
        28.84 & 0.9258 & 	
        28.89 & \textbf{0.9333} &
        35.34 & 0.9695 \\
    \midrule
    
    \textbf{ReRAW-R} & 
        \underline{40.12} & \textbf{0.9915} & 
        \underline{38.64} & \underline{0.9929} & 
        \textbf{30.52} & \textbf{0.9492} & 
        \underline{29.85} & 0.9103 & 
        \underline{38.51} & \underline{0.9860} \\
    \textbf{ReRAW-S} &  
        \textbf{41.00} & \underline{0.9914}& 
        \textbf{40.07} & \textbf{0.9931} & 
        \underline{30.18} &	\underline{0.9466} &	
        \textbf{30.45} & \underline{0.9122} &	
        \textbf{38.88} & \textbf{0.9861} \\
    \bottomrule
  \end{tabular}
  \caption{RGB to RAW reconstruction performance comparison by PSNR (dB) ($\uparrow$) and SSIM($\uparrow$). ReRAW, particularly the stratified sampling variant ReRAW-S, outperforms competing methods. Best result is highlighted in bold, second best underlined.}
  \label{tab:reraw}
\end{table*}

%% file: tables/table-ablation-loss.tex
\begin{table}[!t]
  \centering
  \begin{tabular}{c|ccccc}
    \toprule
        Loss &
        \makecell{\textbf{NOD} \\ \textbf{Nikon}} & 
        \makecell{\textbf{NOD} \\ \textbf{Sony}} & 
        \makecell{\textbf{FIVEK} \\ \textbf{Nikon}} &
        \makecell{\textbf{FIVEK} \\ \textbf{Canon}} & 
        \makecell{\textbf{PASCAL} \\ \textbf{RAW}} \\ 
    \midrule
       $\mathcal{L}_{1}$ & 40.71 & 39.97 & \textbf{31.35} & 29.85 & 38.87 \\ 
       $\mathcal{L}_{2}$ & 40.65 & 39.43 & 28.30 & 27.93 & 38.34 \\ 
       $\mathcal{L}_{hln}$ & \textbf{41.00} & \textbf{40.07} & 30.18 & \textbf{30.45} & \textbf{38.88} \\
    \bottomrule
  \end{tabular}
  \caption{Comparison on different loss function and their effect of ReRAW reverse ISP conversion PSNR (dB) ($\uparrow$). Our proposed $\mathcal{L}_{hln}$ loss achieves tops performance on the majority of datasets.}
  \label{tab:ablation-loss}
  \vspace{-0.2cm}
\end{table}

%% file: tables/table-ablation-architecture.tex
\begin{table}[t]
  \centering
  \begin{tabular}{ccc|cc}
    \toprule
    \makecell{Global \\ Context \\ Encoder} & \makecell{Gamma \\ Predictor \\ no. heads} & 
    \makecell{Gamma \\ Scaling \\ Encoder} 
    & \makecell{\textbf{NOD} \\ \textbf{Nikon}} & 
    \makecell{\textbf{FIVEK} \\ \textbf{Canon}} \\
    \midrule
       & 1 & & 36.04 &	26.55\\ 
     \checkmark & 1 & & 40.95 & 29.5\\ 
      & 2 & & 36.10 & 27.44\\
     \checkmark & 2 & & 40.98 & 29.27\\
     \checkmark & 2 & \checkmark & 40.87 & 29.75\\
     \checkmark & 10 & \checkmark & \textbf{41.00} &	\textbf{30.45}\\     
    \bottomrule
  \end{tabular}
  \caption{Ablation study on the components of ReRAW and their effect on conversion PSNR (dB) ($\uparrow$). Global Context Encoder's impact is high, while both increasing the no. of heads and adding the scaling encoder further boosts performance. When (M) has multiple heads and (G) is not used, the outputs are just averaged.}
  \label{tab:ablation-architecture}
  \vspace{-0.3cm}
\end{table}

%% file: tables/table-mAP.tex
\begin{table*}[t]
  \centering
  \begin{tabular}{l|l|l|ccc|ccc}
    \toprule
    \multirow{2}{*}{Model} & \multirow{2}{*}{Pretraining} & \multirow{2}{*}{Finetuning} &
        \multicolumn{3}{c}{\textbf{PASCALRAW}} & 
        \multicolumn{3}{c}{\textbf{NOD - Nikon}} \\
    & & & mAP & mAP50 & mAP75 & mAP & mAP50 & mAP75 \\

    \midrule
    \multirow{2}{*}{RTMDet-s \cite{rtmdet}} 
    & a) BDD-RGB & GT-RGB &     57.62 & 86.60 & 60.69 & 20.46 & \textbf{39.74} & 18.39 \\
    & b) BDD-RGB & GT-RAW &     56.20 & 85.65 & 58.33 & 20.21 & 38.45 & 18.91 \\
    & c) BDD-ReRAW-R & GT-RAW &  \underline{62.44} & \textbf{91.12} & \underline{65.23} & \textbf{21.53} & \underline{38.64} & \textbf{20.76} \\
    & d) \textbf{BDD-ReRAW-S} & \textbf{GT-RAW} & \textbf{63.19} & \underline{90.45} & \textbf{66.14} & \underline{21.09} & 38.14 & \underline{20.49} \\
    
    \midrule
    \multirow{2}{*}{YOLOX-s \cite{yolox}} 
    & a) BDD-RGB & GT-RGB &     \underline{65.36} & \underline{91.45} & \underline{71.70} & 27.14 & 49.58 & 25.45 \\
    & b) BDD-RGB & GT-RAW &     64.00 & 90.33 & 70.06 & \underline{27.30} & 49.76 & \underline{26.13} \\
    & c) BDD-ReRAW-R & GT-RAW & 62.64 & 90.88 & 67.75 & 27.09 & \underline{50.52} & 25.49 \\
    & d) \textbf{BDD-ReRAW-S} & \textbf{GT-RAW} & \textbf{65.85} & \textbf{91.76} & \textbf{71.73} & \textbf{29.03} & \textbf{52.92} & \textbf{27.49}\\
        
    \midrule
    \multirow{2}{*}{SSD \cite{MobileNet}} 
    & a) BDD-RGB & GT-RGB &     \underline{62.50} & \underline{90.63} & \underline{66.38} & 22.97 & 40.98 & 21.83 \\
    & b) BDD-RGB & GT-RAW &     62.22 & 90.53 & 65.67 & \underline{23.06} & \underline{41.05} & \textbf{22.64} \\
    & c) BDD-ReRAW-R & GT-RAW & 60.96 & 89.60 & 64.53 & 22.56 & 40.28 & 22.30 \\
    & d) \textbf{BDD-ReRAW-S} & \textbf{GT-RAW}  & \textbf{63.09} & \textbf{90.98} & \textbf{67.38} & \textbf{23.40} & \textbf{41.39} & \underline{22.32} \\
    \bottomrule
  \end{tabular}
  \caption{Object detection training results: 3 OD models $\times$ 4 training variants. Training heuristic d), involving pretraining on a mix high quality synthetic RAW dataset converted by ReRAW and ground truth RAW data (BDD-ReRAW-S), then finetuning on a RAW dataset of interest, generally achieves the highest performance in terms of mAP, compared to other training heuristics, including a full RGB pipeline.}
  \label{tab:od}
  \vspace{-0.2cm}
\end{table*}

%% file: sections/5_Discussion.tex
\section{Discussion and Future Work}
\label{sec:discussion}

Our proposed stratified sampling technique helps in boosting the conversion performance of ReRAW, and proved that training data curation is beneficial even for low-level image tasks such as RGB to RAW conversion. We plan to explore other sampling methods such as filtering training patches based on high dynamic range, or high entropy, that might further boost conversion performance.

The experiments in Table \ref{tab:od} showed that maximizing object detection accuracy of small detectors on the edge operating on unprocessed RAW image signal directly, the training recipe listed in d) is the most salient. 
The RTMDet detector showed the highest sensitivity to input domain, where training on PASCALRAW yielded a large variability in mAP results, whilst YOLOX and SSD showed less variability. This suggests that developing detector architectures specifically tailored for RAW images is a promising direction for future research. We also acknowledge that the RTMDet results on NOD-Nikon show preference to the lower PSNR synthetic RAW dataset (BDD-ReRAW-R), which is surprising. This shows a limitation of the PSNR metric when relating synthetic RAW conversion performance and downstream task accuracy, and it's worth exploring further.

%% file: sections/6_Conclusion.tex
\section{Conclusion}
\label{sec:conclusion}

We introduced ReRAW, a state-of-the-art high-PSNR reverse ISP for converting RGB images into sensor-specific RAW. ReRAW achieves the highest reconstruction accuracy against competing state-of-the-art methods, for five different datasets, due to its unique multi-head architecture predicting RAW image candidates in gamma space, and stratified training data sampling technique. Using ReRAW to generate high-quality synthetic RAW datasets for pretraining OD models and fine-tuning on real RAW data results in superior performance compared to models trained on traditional RGB pipelines. This method, thanks to ReRAW’s high reconstruction accuracy, optimizes model training for edge devices, bypassing the ISP hence saving energy and time, and enhancing OD accuracy over standard RGB workflows.